\documentclass[aps,twocolumn,showpacs,amssymb,
floatfix,superscriptaddress]{revtex4}
\usepackage{graphicx}
\usepackage{epsf}
\usepackage{dcolumn}
\usepackage{bm}
\usepackage{hyperref} 
\usepackage{latexsym}

\begin{document}
\def\av#1{\langle#1\rangle}
\def\a{\alpha}
\def\etal{{\it et al.}}
\def\pc{p_{\rm c}}
\def\l{{\lambda}}


\title{Efficient Immunization Strategies for 
Computer Networks and Populations} 
\author{Reuven~Cohen
\footnote  {{\bf e-mail:} cohenr@shoshi.ph.biu.ac.il}}
\author{Shlomo~Havlin} 
\affiliation{Minerva Center and Department of Physics, Bar-Ilan University,
Ramat-Gan, 52900, Israel}
\author{Daniel~ben-Avraham}
\affiliation{Department of Physics,Clarkson University,
Potsdam NY 13699-5820,
USA}

\begin{abstract}
We present an effective immunization strategy for 
computer networks and populations with broad and,
in particular, scale-free degree distributions. 
The proposed strategy, {\it acquaintance immunization}, calls for the
immunization of random acquaintances of random nodes (individuals).
The strategy requires no knowledge
of the node degrees or any other global knowledge, as do targeted
immunization strategies. We study analytically the critical threshold 
for complete immunization. 
We also study the strategy with respect to the SIR 
(susceptible-infected-removed) epidemiological model. We show that 
the immunization threshold is dramatically reduced with 
the suggested strategy, for all studied cases.
\end{abstract}
\maketitle
It is well established that random immunization requires immunizing a
very large fraction of a computer network, or population, in order to
arrest epidemics that spread upon contact between infected nodes (or
individuals)~\cite{AM92,MA84,HV86,WY84,AJB00,cohen,PV01}.  Many
diseases require 80\%-100\% immunization (for example, Measles
requires 95\% of the population to be immunized~\cite{AM92}). The same
is correct for the Internet, where stopping computer viruses requires
almost 100\% immunization~\cite{AJB00,cohen,PV01}.  On the other hand,
targeted immunization of the most highly connected
individuals~\cite{AM92,AJB00,PV01b,LM01,cal,cohen2}, while effective,
requires global information about the network in question, rendering
it impractical in many cases. Here, we develop a mathematical model
and propose an effective strategy, based on the immunization of a
small fraction of {\it random acquaintances} of randomly selected
nodes. In this way, the most highly connected nodes are immunized, and
the process prevents epidemics with a small finite immunization
threshold and without requiring specific knowledge of the network.

\begin{figure}
\includegraphics[width=0.37\textwidth,angle=-90]{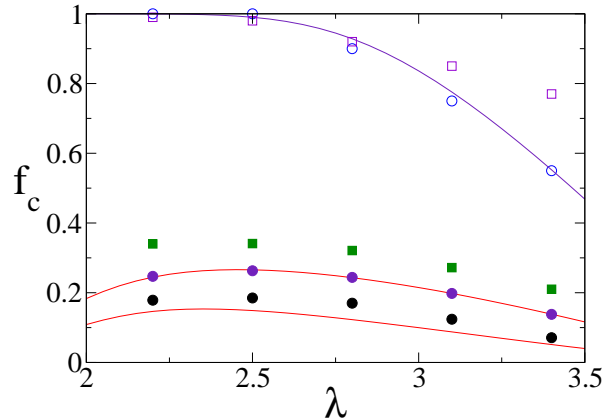}
\caption{
Critical probability, $f_c$, as a function of $\l$ in scale-free
networks (with $m=1$), for the random immunization (top curve and open
circles), acquaintance immunization (middle curve and top full
circles) and double acquaintance immunization (bottom curve and bottom
full circles) strategies. Curves represent analytical results (an
approximate one for double-acquaintance), while data points represent
simulation data, for a population $N=10^6$ [Due to the population's
finite size, $f_c<1$ for random immunization even when $\l<3$].  Squares
are for random (open) and acquaintance immunization (full) of
assortatively mixed networks (where links between sites of degree
$k_1$ and $k_2 (>k_1)$ are rejected with probability
$0.7\left(1-\frac{k_1}{k_2}\right)$).
\label{pc} }
\vskip -0.15in
\end{figure}

Social networks are known to possess a broad distribution of the
number of links (contacts), $k$, emanating from a node (an
individual)~\cite{strogatz,bar_rev,dor_rev}. Examples are the web of
sexual contacts~\cite{lil}, movie-actor networks, science citations
and cooperation networks~\cite{bar_collab,new_pnas} etc.  Computer
networks, both physical (such as the Internet~\cite{internet}) and
logical (such as the WWW~\cite{WWW}, and e-mail~\cite{bornholdt} and
trust networks~\cite{amaral}) are also known to posses wide,
scale-free, distributions.  Studies of percolation on broad-scale
networks show that a large fraction $f_c$ of the nodes need to be
removed (immunized) before the integrity of the network is
compromised.  This is particularly true for scale-free networks,
$P(k)=ck^{-\l}$ ($k\geq m$), where $2<\l<3$ --- the case of most known
networks~\cite{strogatz,bar_rev,dor_rev} --- where the percolation
threshold $f_c\to 1$, and the network remains connected (contagious)
even after removal of most of its nodes~\cite{cohen}.  In other words,
with a random immunization strategy almost all of the nodes need to be
immunized before an epidemic is arrested (see Fig.~\ref{pc}).

When the most highly connected nodes are targeted first, removal of
just a small fraction of the nodes results in the network's
disintegration~\cite{AJB00,cal,cohen2}.  This has led to the
suggestion of targeted immunization of the HUBs (the most highly
connected nodes in the network)~\cite{PV01b,dezso}.  However, this
approach requires a complete, or at least fairly good knowledge of the
degree of each node in the network.  Such global information often
proves hard to gather, and may not even be well-defined (as in social
networks, where the number of social relations depends on subjective
judging).  The acquaintance immunization strategy proposed herein
works at low immunization rates, $f$, and obviates the need for global
information.

In our approach, we choose a random fraction $p$ of the $N$ nodes and
look for a random acquaintance with whom they are in contact (thus,
the strategy is purely local, requiring minimal information about
randomly selected nodes and their immediate environs ). The
acquaintances, rather than the originally chosen nodes, are the ones
immunized.  The fraction $p$ may be larger than $1$~\cite{note}, 
for a node might
be queried more than once, on average, while the fraction of nodes
immunized $f$ is always less than or equal to $1$.

Suppose we apply the acquaintance strategy on a random fraction $p$ of
the network. The critical fractions, $p_c$ and $f_c$, needed to stop
the epidemic can be analytically calculated.  In each event, the
probability that a node with $k$ contacts is selected for immunization
is $kP(k)/(N\av{k})$~\cite{cal,cohen}, where $\av{k}=\sum_k kP(k)$
denotes the average degree of nodes in the network. This quantifies
the known fact that randomly selected acquaintances possess more links
than randomly selected nodes~\cite{feld,ego}.  Suppose we follow some
branch, starting from a random link of the spanning cluster. In some
layer, $l$, we have $n_l(k)$ nodes of degree $k$. In the next layer
($l+1$) each of those nodes has $k-1$ new neighbors (excluding the one
through which we arrived).  Let us denote the event that a node of
degree $k$ is susceptible to the disease (not immunized) by $s_k$.  To
find out the number of nodes, $n_{l+1}(k')$, of degree $k'$ that are
susceptible, we multiply the number of links going out of the $l$th
layer by the probability of reaching a node of degree $k'$ by
following a link from a {\em susceptible} node, $p(k'|k, s_k)$.  Then,
we multiply by the probability that this node is also susceptible
given both the node and the neighbor's degrees, and the fact that the
neighbor is also susceptible, $p(s_{k'}|k', k , s_k)$.  Since below
and at the critical percolation threshold loops are
irrelevant~\cite{cohen}, one can ignore them. Therefore,
\begin{equation}
\label{nl}
n_{l+1}(k')=\sum_k n_l(k)(k-1)p(k'|k,s_k)p(s_{k'}|k',k,s_k)\;.
\end{equation}
By using Bayes' rule:
\begin{equation}
p(k'|k, s_k)=\frac{p(s_k|k, k')p(k'|k)}{p(s_k|k)}\;.
\end{equation}
Assuming that the network is uncorrelated (no degree-degree
correlations), the probability of reaching a node with degree $k'$ via
a link, $\phi(k')\equiv p(k'|k)=k'P(k')/\av{k}$, is independent of
$k$.

A random site (of degree $k$) is selected in each step with
probability $1/N$.  The probability of being redirected to a specific
acquaintance is $1/k$.  Thus, the probability that the acquaintance is
{\it not} selected in one particular attempt, is $(1-1/Nk)$, and in
all $Np$ vaccination attempts, it is
\begin{equation}
\nu_p(k)\equiv \left(1-\frac{1}{Nk}\right)^{Np}\approx e^{-p/k}\;.
\end{equation}
If the neighbor's degree is not known, the probability is
$\nu_p\equiv\av{\nu_p(k)}$, where the average (and all averages
henceforth) is taken with respect to the probability distribution
$\phi(k)$.  The probability that a node with degree $k'$ is
susceptible is $p(s_{k'}|k')=\av{\exp(-p/k)}^{k'}$, if no other
information exists on its neighbors. If the degree of one neighbor is
known to be $k'$: $p(s_k|k, k')=e^{-p/k'}\times\av{e^{-p/k}}^{k-1}$.
Since the fact that a neighbor with known degree is immunized does not
provide any further information about a node's probability of
immunization, it follows that $p(s_k|k, k')=p(s_k|k, k', s_{k'})$.
Using the above equations one obtains:
\begin{equation}
p(k'|k, s_k)=\frac{\phi(k')e^{-p/k'}}{\av{e^{-p/k}}}\;.
\end{equation}

Substituting these results in (\ref{nl}) yields:
\begin{equation}
\label{sum1}
n_{l+1}(k')=\nu_p^{k'-2}\phi(k')e^{-p/k'}\sum_k n_l(k)(k-1)e^{-p/k}\;.
\end{equation}
Since the sum in (\ref{sum1}) does not depend on $k'$, it leads to the
stable distribution of degree in a layer $l$: $n_l(k)=a_l
\nu_p^{k-2}\phi(k)e^{-p/k}$, for some $a_l$. Substituting this into
(\ref{sum1}) yields:
\begin{equation}
n_{l+1}(k')=n_l(k')\sum_k \phi(k)(k-1)\nu_p^{k-2}e^{-2p/k}\;.
\end{equation}
Therefore, if the sum is larger than $1$ the branching process will
continue forever (the percolating phase), while if it is smaller than
$1$ immunization is sub-critical and the epidemic is arrested. Thus,
we obtain a relation for $p_c$:
\begin{equation}
\label{p_c}
\sum_k \frac{P(k)k(k-1)}{\av{k}}\nu_{p_c}^{k-2}e^{-2p_c/k}=1\;.
\end{equation}

The fraction of immunized nodes is easily obtained from the fraction
of nodes which are not susceptible,
\begin{equation}
\label{f_c}
f_c=1-\sum_k P(k)p(s_k|k)=1-\sum_k P(k)\nu_{p_c}^k\;,
\end{equation}
where $P(k)$ is the regular distribution, and $p_c$ is found
numerically using Eq. (\ref{p_c}).

A related immunization strategy calls for the immunization of
acquaintances referred to by at least $n$ nodes. (Above, we
specialized to $n=1$.)  The threshold is lower the larger $n$ is, and
may justify, under certain circumstances, this somewhat more involved
protocol.

The acquaintance immunization strategy is effective for any
broad-scale distributed network. Here we give examples for scale-free
and bimodal distributions, which are common in many natural networks.
We also give an example of an assortatively mixed network (where high
degree nodes tend to connect to other high degree
nodes~\cite{assort}).  We also discuss the effectiveness of the
strategy in conjunction with the SIR epidemiological model.

In Fig.~\ref{pc}, we show the immunization threshold $f_c$ needed to
stop an epidemic in networks with $2<\l<3.5$ (this covers all known
cases).  Plotted are curves for the (inefficient) random strategy, and
the strategy advanced here, for the cases of $n=1$ and $2$. Note that
while $f_c=1$ for networks with $2<\l<3$ (e.g. the Internet) it
decreases dramatically to values $f_c\approx 0.25$ with the suggested
strategy. The figure also shows the strategy's effectiveness in case
of assortatively mixed networks~\cite{assort}, {\it i.e.}, in cases
where $p(k'|k)$ does depend on $k$, and high degree nodes tend to
connect to other high degree nodes, which is the case for many real
networks.

\begin{figure}
\includegraphics[width=0.37\textwidth,angle=-90]{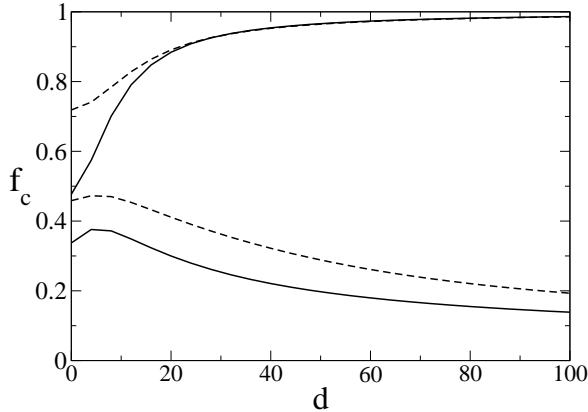}
\caption{
Critical concentration, $f_c$, for the bimodal distribution (of two
Gaussians) as a function of $d$, the distance between the modes. The
first Gaussian is centered at $k=3$ and the second one at $k=d+3$ with
height 5\% of the first. Both have variance $2$ (solid lines) or $8$
(dashed lines). Top 2 lines are for random immunization. The bottom 2
lines are for acquaintance immunization.  All curves are analytically
derived from Eqs. (\ref{f_c}) and (\ref{p_c}).  Very similar results
have been obtained for bimodal distributions of two Poissonians. Note
that also for the case $d=0$, i.e. a single Gaussian, the value of
$f_c$ reduces considerably due to the acquaintance immunization
strategy. Thus the strategy gives improved performance even for
relatively narrow distributions~\cite{amaral2}.
\label{pc_gauss} }
\vskip -0.15in
\end{figure}

\begin{figure}
\vskip 0.13in
\includegraphics[width=0.37\textwidth]{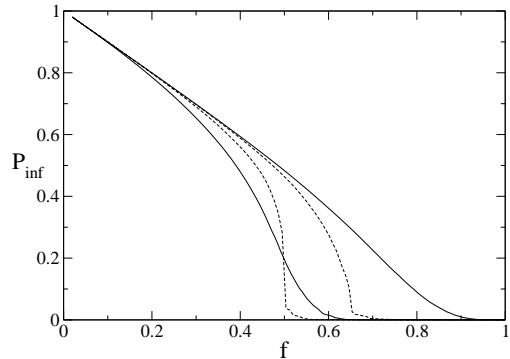}
\caption{The fraction of the population infected in the endemic state,
$P_{inf}$, as a function of $f$, the immunized fraction of the
population, for a scale-free network embedded in a $2d$ geographical
space (see~\cite{RCBH02}).  The solid lines are for random (top) and
acquaintance (bottom) immunization for a network with $\l=2.8$ and the
dashed lines are for the same cases with $\l=4$.  In all cases
$N=10^6$ and $m=4$.  In both cases ($\l=2.8$ and $4$) the acquaintance
immunization strategy provides a considerable improvement over random
immunization. The high values for $f_c$ stem from the fact that the
network is very well connected with $m=4$, which was taken in order to
approach a regular square lattice at $\l\to\infty$
\label{geog}}
\vskip -0.15in
\end{figure}

Fig.~\ref{pc_gauss} gives similar results for a bimodal distribution
(consisting of two Gaussians, where high degree nodes are rare
compared to low degree ones). This distribution is also believed to
exist for some social networks, in particular, for some networks of
sexual contacts. In Fig.~\ref{geog} geographical effects, where nodes
tend to connect to geographically adjacent ones~\cite{RCBH02}, are
also taken into account. The improvement gained by the use of the
acquaintance immunization strategy is evident in both cases, as seen
in Figs. \ref{pc_gauss} and \ref{geog}.

\begin{figure}
\includegraphics[width=0.37\textwidth,angle=-90]{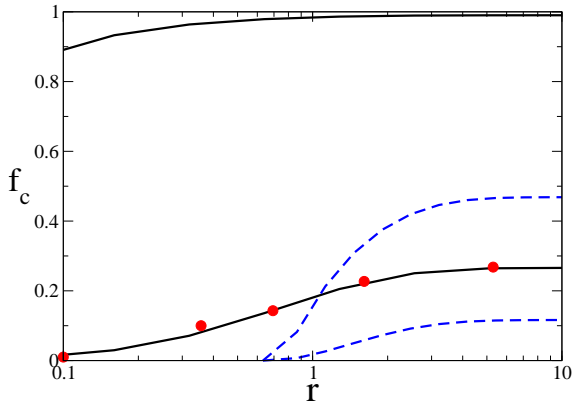}
\vskip 0.1in
\caption{ Critical concentration, $f_c$, vs $r$, the infection rate,
for the SIR model with $\tau=1$. The solid lines are for random (top)
and acquaintance immunization (bottom) for scale-free networks with
$\l=2.5$. The dashed lines are for $\l=3.5$ (top -- random, bottom --
acquaintance immunization).The circles represent simulation results
for acquaintance immunization for scale-free networks with $\l=2.5$.
\label{pc_sir} }
\vskip -0.15in
\end{figure}

The above considerations hold if full immunization is required. That
is, given a static network structure, one wishes to stop any epidemic
or virus propagation. However, most real viruses have a finite
infection rate, and, therefore, a finite probability of infecting a
neighbor of an infected node.  The SIR model, widely studied by
epidemiologists~\cite{grass,sok,new}, assumes that nodes can be
susceptible, infected, or removed ({\it i.e.}  recovered and immunized
against further infection or otherwise removed from the network). This
epidemiological model can be mapped to a bond percolation model, where
the concentration of bonds, $q=1-e^{-r\tau}$, where $r$ is the
transmissibility of the virus (infection rate over a link) and $\tau$
is the infection time. To find the effect of the strategy, given this
finite infection probability, the right hand side of Eq. (\ref{nl})
should be multiplied by $q$, giving:
\begin{equation}
\sum_k \frac{P(k)k(k-1)}{\av{k}}\nu_{p_c}^{k-2}e^{-2p_c/k}=q^{-1}\;.
\end{equation}
instead of Eq. (\ref{p_c}). Results for different infection rates and
scale-free networks with $\l=2.5$ and $\l=3.5$ are shown in
Fig.~\ref{pc_sir}. As can be seen in the figure, in the limit
$\tau r\to\infty$ this model leads to the full immunization case of
Fig.~1. For lower values of $r$, the proposed strategy still gives
similar, or even larger, decrease in the immunization threshold.

Various immunization strategies have been proposed, mainly for the
case of an already spread disease, and are based on tracing the chain
of infection towards the super-spreaders of the disease~\cite{WY84}.
This approach is different from our proposed approach, since it is
mainly aimed at stopping an epidemic after the outbreak began. It is
also applicable for cases where no immunization exists and only
treatment for already infected individuals is possible.  Our approach,
on the other hand, can be used even before the epidemic starts
spreading, since it does not require any knowledge of the chain of
infection.

In practice, any population immunization strategy must take into
account issues of attempted manipulation. We would expect the
suggested strategy to be less sensitive to manipulations than targeted
immunization strategies. This is due to its dependence on acquaintance
reports, rather than on {\it self}-estimates of number of contacts.
Since a node's reported contacts pose a direct threat to the node (and
relations), we anticipate that manipulations would be less
frequent. Furthermore, we would suggest adding some randomness to the
process: for example, reported acquaintances are not immunized, with
some small probability (smaller than the random epidemic threshold),
while randomly selected individuals are immunized directly, with some
low probability.  This will have a small impact on the efficiency,
while enhancing privacy and rendering manipulations less practical.

In conclusion, we have proposed a novel efficient strategy for
immunization, requiring no knowledge of the nodes' degrees or any
other global information. This strategy is efficient for networks of
any broad-degree distribution and allows for a low threshold of
immunization, even where random immunization requires the entire
population to be immunized.  We have presented analytical results for
the critical immunization fraction in both a static model and the
kinetic SIR model.

As a final remark, we note that our approach may be relevant to other
networks, such as ecological networks of
predator-prey~\cite{eco,eco2}, metabolic networks~\cite{metabolic},
networks of cellular proteins~\cite{cell}, and terrorist networks.
For terrorist networks, our findings suggest that an efficient way to
disintegrate the network, is to focus more on removing individuals
whose name is obtained from another member of the network.

\acknowledgments
We are grateful to NSF grant PHY-0140094 (DbA) for partial support 
of this research.


\begin{thebibliography}{99}
\bibitem{note}
   For most cases $p_c$ is well below $1$. For the scale free distribution
   for $\lambda=2.5$, $3$ and $3.5$, $p_c\approx 0.81$, $0.52$ and $0.28$
   respectively. Querying  is, however, not expensive, and asking each
   individual for more than one neighbor will decrease these numbers
   significantly, to approx. $0.4$, $0.3$ and $0.2$, respectively.
\end{thebibliography}
\end{document}